# Proximity-induced superconductivity effect in a double-stranded DNA

Hamidreza Simchi[*,1,2], Mahdi Esmaeilzadeh[**,1], and Hossein Mazidabadi[1]

[1]Department of Physics, Iran University of Science and Technology, Narrmak, Tehran 16844, Iran

[2]Semiconductor Technology Center, Tehran Iran

We study the proximity-induced superconductivity effect in a double-stranded DNA by solving the Bogoliubov-de Gennes equations and taking into account the effect of thermal fluctuations of the twist angle between neighboring base pairs. We show that the electron conductance is spin-dependent and the conductance of spin up (down) increases (decreases) due to the spin-orbit coupling. It is found that, for T<100K, the band gap energy is temperature-independent and it decreases due to the SOC. In addition, by solving the Bogoliubov-de Gennes equations and local gap parameter equation self-consistently, we find the critical temperature at which transition to superconductivity can take place.

* E-mail: simchi@iust.ac.ir,
** E-mail: mahdi@iust.ac.ir

I. Introduction

The self-assembly property, the ability to synthesize in different sequences, and the highly specific binding between the single-stranded DNA have made it a suitable candidate for use in molecular electronics.[1] It has been shown that the DNA has metallic[2-5], semiconducting[6-10], and insulating[11-12] behaviors. Kasumov *et al.* have found that the proximity-induced superconductivity in DNA molecules can occur below the superconducting transition temperature of the electrodes.[13] As shown by Ren *et al.*[14], thermal fluctuation of twist-angle between the neighboring base pairs causes the averaged hopping matrix elements to decrease and fluctuate which leads to a significant thermal enhancement of charge transport at low temperatures when the DNA is in contact with superconducting leads. Gohler *et al.* have reported the spin-selective transmission property of electrons through the self-assembled monolayer of double-stranded DNA (dsDNA) on gold.[15] Guo *et al.*[16] have shown that the combination of the spin orbit coupling (SOC), the environment-induced dephasing, and the helical symmetry could be considered for describing the experimentally studied system by Gohler *et al.*[15] It has been shown that, based on Hamiltonian introduced by Guo *et al.*[16,17], dsDNA can act as a field effect transistor.[18]

The induced proximity effect in DNA[13], the thermal enhancement of charge transport[14], and the spin-selective transmission of electrons through the self-assembled monolayer of dsDNA on gold,[15,16] motivated us to study the effect of temperature on the spin transport properties in dsDNA. We assume that the twist-angle of the neighboring base pairs obeys a Gaussian distribution such that its mean value and variance are zero and T/250, respectively,[14] where T is the temperature in Kelvin. Also, it is assumed that $tcos(\theta)$ is the temperature dependent hopping matrix element, where $t$ is the hopping matrix element at zero temperature and $\theta$ is the randomly fluctuating twist-angle between the neighboring base pairs. Using Guo's Hamiltonian[16] and the non-equilibrium Green's function method; we calculate the conductance of dsDNA at different values of temperature. We show that the conductance of dsDNA is spin-dependent and the conductance of spin up increases due to the SOC, while the conductance of spin down decreases. It is shown that, for T<100K, the band gap energy is constant and temperature-independent. Also, we show that the spin-orbit coupling decreases the band gap energy for this region of temperature. In addition, to study the proximity-induced superconductivity effect in dsDNA and to determine the local gap parameter and critical temperature, we solve self-consistently the single-electron Hamiltonian of Bogoliubov-de Gennes (BdG) equations including environment-induced dephasing and the local gap parameter equation. It is shown that the critical temperature is 0.25 K which in agreement with experimental data.[13] The organization of this paper is as follows. In Section II, we present the calculation method. The results and discussion are presented in Section III. Section IV contains a summary.

## II. Calculation method

The BdG equations can be written as[19-21]

$$\begin{pmatrix} H & \Delta \\ \Delta^* & -H \end{pmatrix} \begin{pmatrix} u_m \\ v_m \end{pmatrix} = \varepsilon_m \begin{pmatrix} u_m \\ v_m \end{pmatrix} \quad (1)$$

where $H$ is the single electron Hamiltonian, $\Delta$ is the local gap parameter which can be determined via the self-consistent method, $u_m$ and $v_m$ are the electron eigenfunctions, and $\varepsilon_m$ is the electron energy eigenvalue. The local gap parameter $\Delta$ is given by[19-21]

$$\Delta = g \sum_m u_m v_m [1 - 2f_F(\varepsilon_m)] \tag{2}$$

where $g$ is the coupling constant, $f_F(\varepsilon) = (e^{\varepsilon/(k_B T)} + 1)^{-1}$ is the Fermi function and $k_B$ is the Boltzmann constant. In Eq. (1), the single-electron Hamiltonian H, can be written as[16]

$$\begin{aligned}
H = &\sum_j \left( \sum_n \varepsilon_{ij} c_{jn}^+ c_{jn} + \sum_{n=1}^{N-1} t_j c_{jn}^+ c_{jn+1} + H.c. \right) + \sum_n (\lambda c_{1n}^+ c_{2n} + H.c.) + \\
&\sum_{jn} \{it_{SO} c_{jn}^+ [\sigma_n^{(j)} + \sigma_{n+1}^j] c_{j+1} + H.c.\} + \sum_{jnk} (\varepsilon_{jnk} b_{jnk}^+ b_{jnk} + t_d b_{jnk}^+ b_{jnk} c_{jn} + H.c.) + \\
&\sum_{jk} (t_L a_{Lk}^+ c_{jN} + t_R a_{Rk}^+ c_{jN} + H.c.) + \sum_{k, \beta = L, R} \varepsilon_{\beta k} a_{\beta k}^+ a_{\beta k}
\end{aligned} \tag{3}$$

where $j = 1, 2$ is the strand label and $n \in [1, N]$ denotes nth base-pair of the dsDNA, N is the total number of base pairs, $\varepsilon_{jn}$ is the on-site energy, $t_j$ is the interaction hopping integral, and $\lambda$ is the interchain hybridization interaction. Also, $t_{SO}$ is the SOC strength and $\sigma_{n+1}^j \equiv \sigma_z \cos\theta - (-1)^j [\sigma_x \sin\phi - \sigma_y \cos\phi] \sin\theta$, $\sigma_x, \sigma_y, \sigma_z$ are the Pauli matrices, $\theta$ is the helix angle, $\phi \equiv n\Delta\phi$ is the cylindrical coordinate (azimuth angle), and $\Delta\phi$ is the twist angle which, as mentioned before, is temperature dependent. In Eq. (3), the fourth term is introduced to simulate the phase-breaking process, the fifth term describes the coupling between the leads and dsDNA, and the last term describes the leads.[16-18]

For numerical calculations, we choose the value of system parameters as:

$\varepsilon_{1n}^{(0)} = 0$, $\varepsilon_{2n}^{(0)} = 0.3$, $t_1 = 0.12$, $t_2 = -0.1$, $\lambda = -0.3$, $t_{SO} = 0.01$, $\theta = 0.66$ radian, $\Delta\phi = \pi/5$, $\Gamma_{L,R} = 1$, $\Gamma_d = 0.005$ and N=30. It should be noted that we restrict ourselves to the positive electron energy eigenvalues (i.e., $\varepsilon_m > 0$). Therefore, $f_F(\varepsilon_m)$ in Eq. (2) vanishes at the zero temperature limit.[19-21] The Eq. (1) and Eq. (2) are satisfied simultaneously, therefore we should solve them in self-consistent method. The necessary algorithm for calculating the gap parameter, $\Delta$ is shown in Fig.1. As the figure shows, at first we guess a small value (e.g., 1E-8) for $\Delta$ and then solve Eq. (1). By solving Eq. (1), eigenfunctions and eigenvalues are found. As it is mentioned above, we should only consider the eigenfunctions with positive eigenvalues. Therefore, we sort and choose only these kinds of eigenfunctions. Using Eq. (2), we calculate $\Delta$. Now we should compare the calculated $\Delta$ with guessed $\Delta$. If the difference between calculated $\Delta$ and guessed $\Delta$ is not less than a small value, $\varepsilon$ (e.g., 1E-3), the difference between the $\Delta$s is multiplied to 0.1 and is added to calculated $\Delta$. This is considered as new $\Delta$ for solving the BdG equations.[22] The cycle is repeated unless the difference between $\Delta$s is less than $\varepsilon$. It should be noted that, the gap parameter depends on temperature and in the neighborhood of critical temperature it follows the approximate relation:[23]

$$\Delta(T) \cong 3.2 k_B T_C (1 - \frac{T}{T_C})^{1/2} \qquad (4)$$

where $k_B$ is Boltzmann constant ($8.62 \times 10^{-5}$ eV K$^{-1}$) and $T_C$ is critical temperature. Therefore, for $T \to T_C$ then $\Delta(T) \to 0$. Also, it can be shown that, between $\Delta(T = 0)$ and $T_C$ the below relation is satisfied:[21,23]

$$T_C = \frac{\Delta(T = 0)}{1.76 k_B} \qquad (5)$$

It has been shown that, the experimental values of $2\Delta$ for different materials and different direction in $k$ space generally fall in the range from $3.0 k_B T_C$ to $4.5 k_B T_C$, with most clustered near the Bardeen, Copper, and Schrieffer (BCS) value of $3.5 k_B T_C$.[21] Therefore, by assuming $\Delta(T=0) \equiv \Delta/2$ and using the above algorithm and Eq. 5, we find the $\Delta(T=0)$ and critical temperature which are equal to 0.38 μeV and 0.25 K, respectively. We assume $\Delta = 0.76$ μeV in next calculations.

Also, in our numerical calculations, the electron energy range (i.e., -1 to 1 eV) is divided to 1000 equal parts. Then for each value of energy and for each temperature value in Kelvin, 1000 of twist angles are chosen randomly using Gaussian distribution. Finally, the above obtained averaged values is used to calculate the conductance of dsDNA for each energy.

### III. Results and discussion

Figure 2 shows the conductance of dsDNA versus the electron energy for different values of temperature when the SOC is not considered. It is seen that the electron energy range from -0.45 to -0.1 eV belongs to the occupied molecular orbitals (OMO) and the electron energy range from 0.45 to 0.8 eV belongs to the unoccupied molecular orbitals (UMO). As the figure shows, the difference between the conductance curves for different values of temperature is significant at the edge of OMO and UMO and it decreases with increasing (decreasing) the energy value for UMO (OMO). The band gap energy is defined as the difference between the electron energy of the highest occupied molecular orbital (HOMO) and the energy of the lowest unoccupied molecular orbital (LUMO). As the figure shows, for T<100K (i.e., T = 10, 50, and 100K) the band gap energy is constant which is in agreement with the result of Ref.14.

We now investigate the effect SOC on the conductance of dsDNA. Figure 3 shows the conductance of dsDNA versus the electron energy in the presence of SOC, (a) for spin up and (b) for spin down electrons. By comparing the Fig. 2 and Figs. 3(a) and 3(b), it can be concluded that the conductance of spin up (spin down) electrons increases (decreases) a little due to the SOC. In addition, a comparison between Figs. 3(a) and 3(b) shows that the conductance of spin up electrons and spin down electrons differs considerably from each other at the edge of OMO and UMO. This result is in agreement with the result of Ref. 16. Also, as the Figs. 3(a) and 3(b) show, the band gap energy is constant and spin-independent for T<100K. By attention to the Figs. 1 and 2, it is seen that the bang gap energy is equal to 0.7 eV (0.74 eV) when SOC is (is not) considered for T<100K. Therefore, SOC decreases the band gap energy about 40 meV.

Figures 4 shows the spin-dependent conductance of dsDNA when gap parameter is chosen equal to 0.76 meV. The spin-polarization is defined by $P_s = (G_\uparrow - G_\downarrow)/(G_\uparrow + G_\downarrow)$, where $G_{\uparrow(\downarrow)}$ is the electron conductance with spin up (down). As Fig. 4 shows, the conductance of spin up electrons is completely separated from the conductance of spin down electrons. In the other words, in the regions of energy in which the conductance with spin up is non-zero the conductance with spin down is zero and vice versa. Therefore, the dsDNA can be considered as a prefect spin filter.

The efficiency of a prefect spin-filter depends on its electron conductance value. As shown in Fig. 4, the maximum value of electron conductance for spin down which occurs at E=0.23eV is greater than that for spin up which takes place at E=0.53eV. Therefore, the maximum efficiency of spin down filter is higher than that of spin up filter. In the previous studies[15,16], spin polarization with $P_s = 0.6$ has been found for dsDNA which shows that although spin-filtering can occur, the spin-filtering is not prefect because a considerable fraction (20%) of electrons with opposite spin can also transmit through the dsDNA.

## IV. Summary

In summary, by considering the thermal fluctuation of twist-angle between the neighboring base pairs of dsDNA as a Gaussian distribution, we have shown that, for different temperature, the electron conductance curves differ from each other especially at the edge of OMO and UMO regions. We have also shown that the conductance is spin-dependent and the conductance of spin up (down) electrons increases (decreases) due to the SOC. It has been shown that the bad gap energy is constant and temprature-independent when T<100K. Also, the SOC decreases the band gap energy about 40 meV for this range of temperature. Finally, by solving the BdG-equations and local gap parameter equation self-consistently, we have found the critical temperature and it has been shown that the dsDNA can act as a spin-filter for both spin up and spin down electrons. The spin filtering is prefect and the efficiency of spin down filter is higher than that of spin up.

**References**

[1] R. G. Endres, D. C. Cox, and R. R. P. Singh, Rev. Mod. Phys. **76**, 195 (2004).

[2] Y. Okahata, T. Kobayashi, K. Tanaka, and M. Shimomura, J. Am. Chem. Soc. **120**, 6165 (1998).

[3] H. W. Fink and C. Schonenberger, Nature, **398**, 407 (1999).

[4] A. Rakitin, P. Aich, C. Papadopoulos, Y. Kobzar, A. S. Vedeneev, J. S. Lee, and J. M. Xu, Phys. Rev. Lett. **86**, 3670 (2001).

[5] O. Legrand, D. Cote, and U. Bockelmann, Phys. Rev. E, **73**, 031925 (2006).

[6] D. Porath, A. Bezryadin, S. Vries, and, C. Dekker, Nature **403**, 653 (2003).

[7] K. H. Yoo, D.H. Ha, I.O. Lee, I.W. Park, I. Kim, I.I. Kim, H.Y. Lee, T. Kawai, and H.Y. Choi, Phys. Rev. Lett. **87**, 198102 (2001).

[8] J. S. Hwang, K. J. Kong, D. Ahn, G. S. Lee, D. J. Ahn, and S. W. Hwang, Appl. Phys. Lett. **81**, 1134 (2002).

[9] B. Xu, P. Zhang, X. Li, N. Tao, Nano Lett. **4**, 1105 (2004).

[10] H. Cohen, C. Nogues, R. Naaman, and D. Porath, Proc. Natl. Acad. Sci. **102**, 11589 (2005).

[11] E. Braun, Y. Eichen, U. Sivan, G. Ben-Yoseph, Nature, **391**, 775 (1998).

[12] A. J. Storm, J. van Noort, S. de Vries, and C. Dekker, Appl. Phys. Lett. **79**, 3881 (2001).

[13] A. Yu. Kasumov, M. Kociak, S. Guéron, B. Reulet, V. T. Volkov, D.V. Klinov and H. Bouchiat, Science, **291**, 280 (2001).

[14] W. Ren, J. Wang, Z. Ma, and H. Guo, Phys. Rev. B **72**, 035456 (2005),

[15] B. Gohler, V. Hamelbeck, T. Z. Markus, M. Kettner, G. F. Hanne, Z. Vager, R. Naaman, and H. Zacharias, Science **331**, 894 (2011).

[16] Ai-Min Guo, and Qing-feng Sun, Phys. Rev. Lett. **108**, 218102 (2012).

[17] Ai-Min Guo, and Qing-feng Sun, Phys. Rev. B **86**, 035424 (2012).

[18] H.Simchi, M.Esmaeilzadeh, and H. Mazidabadi, J. Appl. Phys, **113**, 074701 (2013).

[19] K. Halterman and O. T. Valls, arXiv:cond-mat/0107232v1 [cond-mat.supr-con] (2001).

[20] A. A. Shanenko, M. D. Croitoru, and F. M. Peeters, Phys. Rev. B **75**, 014519 (2007).

[21] Michael Tinkham, "*Introduction to superconductivity*", (McGraw-Hil Inc. 1996).

[22] Supriyo Datta, "*Quantum transport: Atom to transistor*", (Cambridge university press, 2005).

[23] T. Van Duzer and C.W. Turner, "*Principles of Superconductive Devices and Circuits*", (Elsevier North Holland Inc. 1981).

**List of Figures**

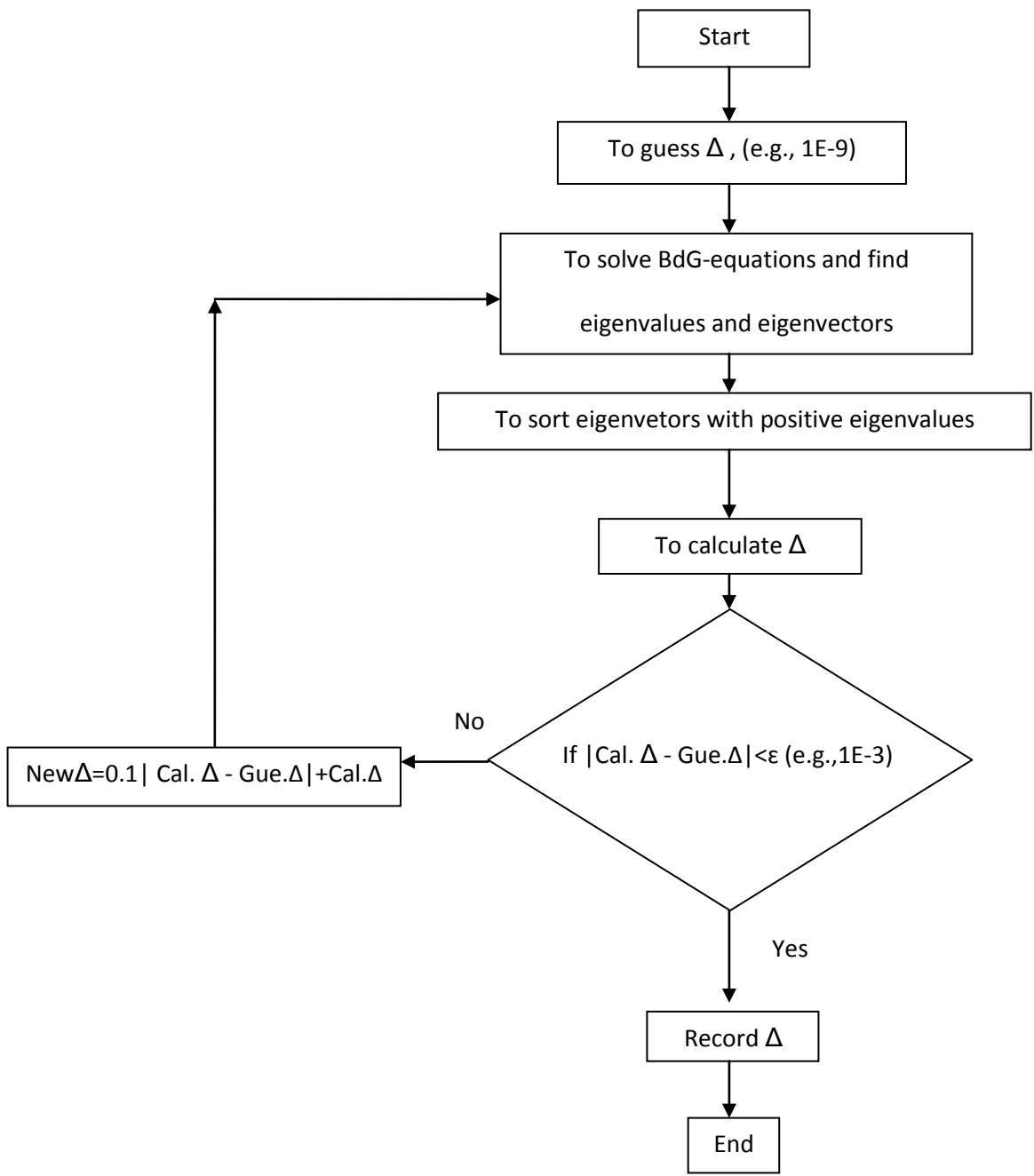

**Fig. 1 Calculation algorithm for calculating the gap parameter using BdG-equations**

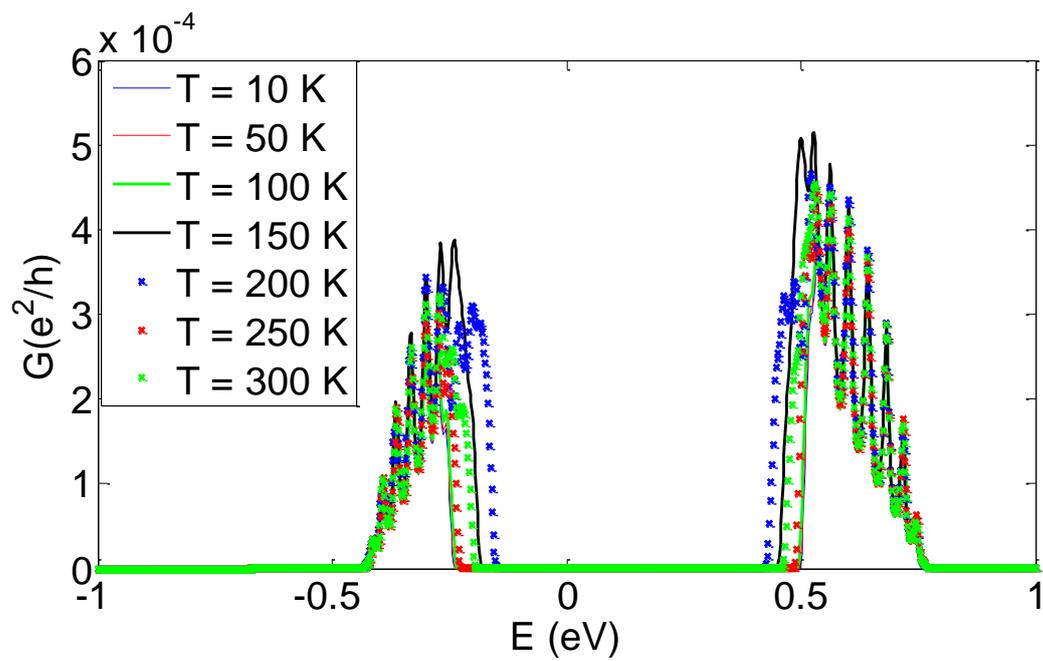

**Fig. 2**

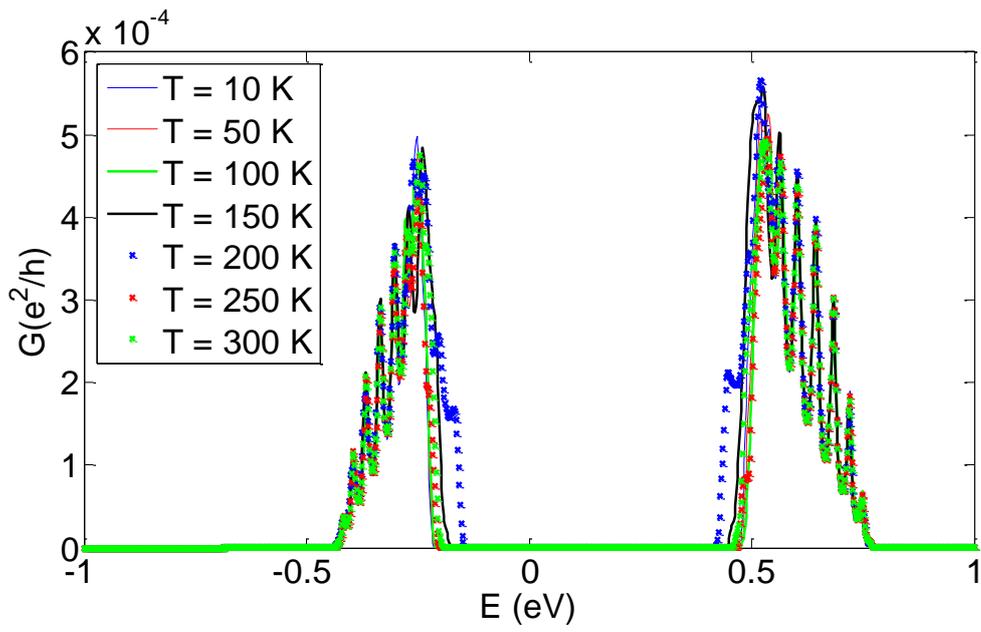

(a)

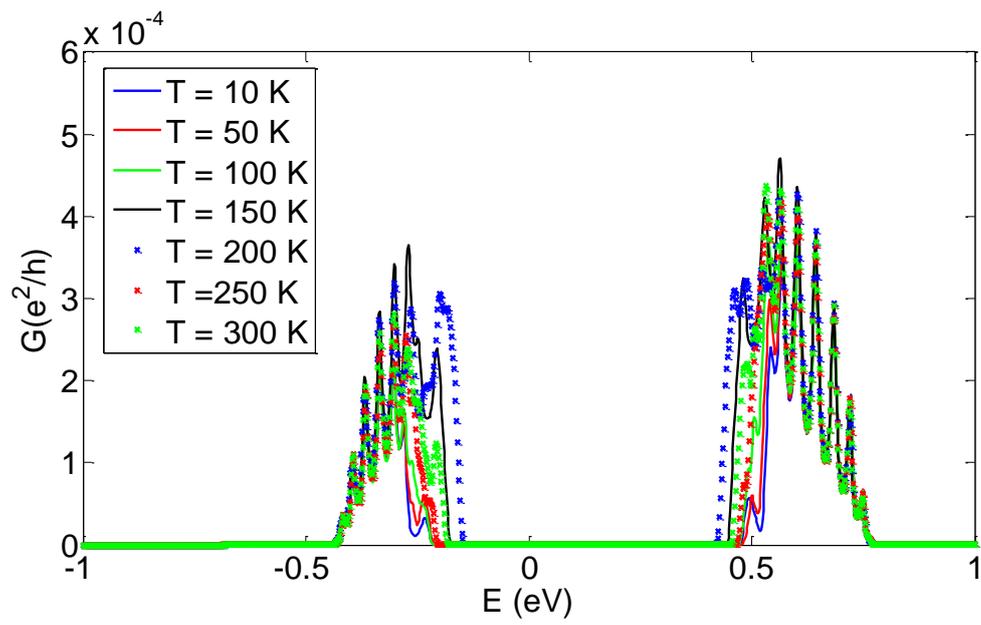

(b)

**Fig. 3**

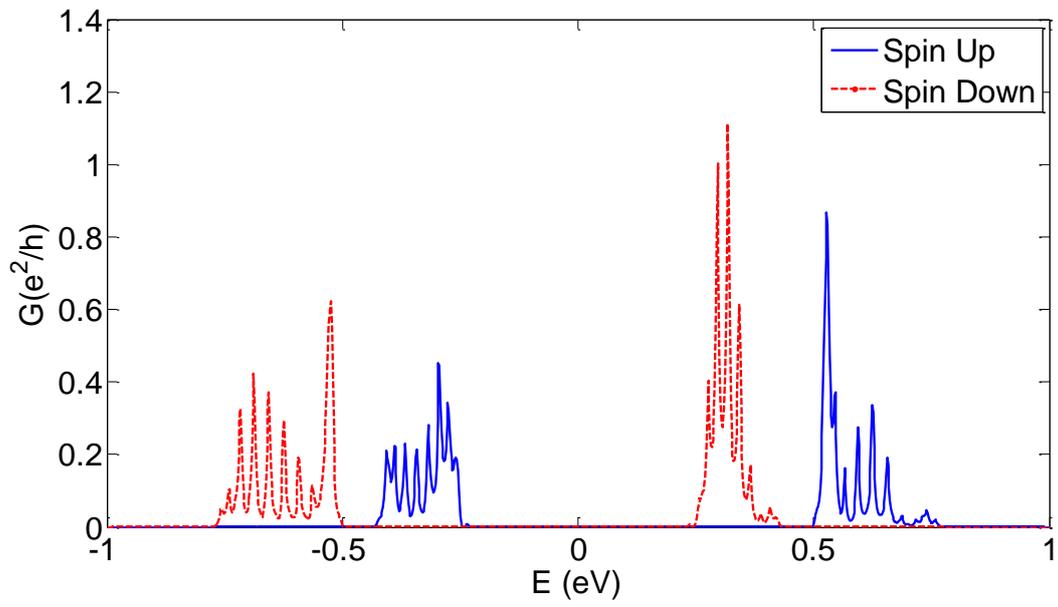

**Fig. 4**